\documentclass[aps,pra,amsfonts,floatfix,superscriptaddress,twocolumn,longbibliography,nofootinbib]{revtex4-1}

\usepackage{graphicx}
\usepackage{dcolumn}
\usepackage{bm}
\usepackage{amsmath}
\usepackage{amssymb}
\usepackage{bbold}
\usepackage{physics}
\usepackage{float}
\usepackage{mathtools}
\usepackage{multirow}
\usepackage{tikz}
\usetikzlibrary{quantikz}
\usepackage{adjustbox}
\usepackage{soul,color}
\usepackage{xcolor}
\usepackage[hidelinks, colorlinks=true, allcolors=blue]{hyperref}

\newcommand{\changed}[1]{\textcolor{black}{#1}}

\begin{document}
\preprint{APS/123-QED}
\title{Simulation of open quantum systems via low-depth convex unitary evolutions}

\author{Joseph Peetz}
 \email{peetz@ucla.edu}
\affiliation{Department of Physics and Astronomy, University of California, Los Angeles, California 90095, USA}

\author{Scott E. Smart}
\affiliation{College of Letters and Science, University of California, Los Angeles, California 90095, USA}

\author{Spyros Tserkis}
\affiliation{College of Letters and Science, University of California, Los Angeles, California 90095, USA}

\author{Prineha Narang}
\email{prineha@ucla.edu}
\affiliation{College of Letters and Science, University of California, Los Angeles, California 90095, USA}

\date{June 10, 2024}

\begin{abstract}
Simulating physical systems on quantum devices is one of the most promising applications of quantum technology. Current quantum approaches to simulating open quantum systems are still practically challenging on NISQ-era devices, because they typically require ancilla qubits and extensive controlled sequences. In this work, we propose a hybrid quantum-classical approach for simulating a class of open system dynamics called random-unitary channels. These channels naturally decompose into a series of convex unitary evolutions, which can then be efficiently sampled and run as independent circuits. The method does not require deep ancilla frameworks and thus can be implemented with lower noise costs. We implement simulations of open quantum systems up to dozens of qubits and with large channel \changed{ranks}.
\end{abstract}

\maketitle

\section{Introduction} \label{section: intro}

System-environment interactions play an important role in the dynamics of quantum systems, giving rise to phenomena such as dissipation, decoherence, relaxation, and particle or energy transfer processes \cite{breuer_theory_2007,Head-Marsden2020}. Ideal simulations of open quantum systems on classical computers suffer from exponentially scaling memory and runtime requirements, leaving complex physical systems largely out of reach without major approximations \cite{brown_using_2010}. Using quantum computers, several proposed techniques for simulating open systems exist, with a common \changed{approach being the encoding of nonunitary dynamics in} a larger, dilated Hilbert space \cite{paulsen_completely_2003}. Numerous techniques make use of Stinespring dilation \cite{stinespring_positive_1955, hu_quantum_2020}, Sz.-Nagy dilation \cite{langer_b_1972,Schlimgen2021}, and linear combination of unitaries \cite{childs_hamiltonian_2012, suri_two-unitary_2022,head-marsden_capturing_2021}. In practice, this dilation leads to long controlled gate sequences and large ancilla frameworks, both of which are practically at odds with NISQ-oriented applications.

\changed{Among a wide variety of open system processes, random-unitary channels form an important and practical subset. Also known as mixed-unitary, these channels} admit an operator-sum decomposition \cite{sudarshan_stochastic_1961, hellwig_operations_1970} of the form
\begin{equation} \label{eq:randomunitary}
\mathcal{E}(\rho)=\sum_i p_i \left(U_i \rho U_i^{\dagger}\right),
\end{equation}
where $p_i$ are probabilities and $U_i$ are unitary operators \cite{mendl_unital_2009}. Random-unitary channels naturally represent many probabilistic processes, including the well-known Pauli channels, and have found widespread utility across quantum information science. In fact, general Markovian processes can be mapped onto effective random-unitary channels through Pauli twirling \cite{dur_standard_2005, cai_constructing_2019} and randomized compiling techniques \cite{wallman_noise_2016}. They are particularly useful for noise modeling of quantum devices \cite{moueddene_realistic_2020, suzuki_qulacs_2021}, as well as within error correction \cite{terhal_quantum_2015} and mitigation \cite{temme_error_2017} schemes. They have also been implemented \changed{in quantum} simulations of thermal relaxation \cite{rost_simulation_2020, tolunayHamiltonianSimulationQuantum2023}. As quantum computers continue to grow in size, efficient simulation of random-unitary channels thus becomes increasingly imperative.

To address this need, we propose a hybrid quantum-classical method for simulating random-unitary channels. Our approach involves classically sampling a channel's probability distribution $\{p_i\}$, and then running a series of corresponding circuits on a quantum computer. Compared to ancilla-based methods, this reduces both the number of qubits and the depth of circuits required, thereby widening the class of simulations feasible on near-term quantum computers. By outlining a practical and efficient algorithm, we seek to enhance both the understanding and utility of random-unitary channels.

\begin{figure*}[t!]
\centering
\includegraphics[width=0.97\textwidth]{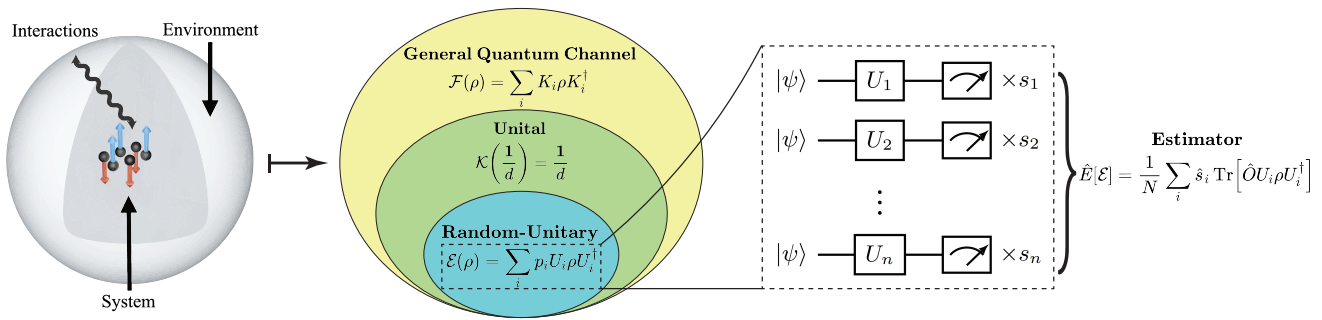} 
\caption{Schematic for simulating random-unitary channels via low-depth convex unitary evolutions. Open system dynamics are described mathematically by their operator-sum representation $\mathcal{F}$, where $K_i$ are Kraus operators. Processes which preserve the maximally mixed state are called unital channels, of which random-unitary channels $\mathcal{E}$ are the subset described by 
\changed{Eq.~\eqref{eq:randomunitary}}. An observable $\hat{O}$ of states evolved via random-unitary channels can be simulated via the low-depth, parallel quantum circuits depicted. Here, each circuit corresponding to the unitary $U_i$ is simulated with $\mathbb{E}[s_i]$ shots, generated from the multinomial distribution $\operatorname{Mult}(N,\{p_i\})$. Each measurement is performed with respect to the observable $\hat{O}$, and the results are then aggregated to compute the unbiased estimator $\hat{E}$.}
\label{big_fig}
\end{figure*} 

\section{Random-Unitary Estimation} \label{section: results}

We present a hybrid quantum-classical method of simulating random-unitary channels for open quantum systems. 
For \changed{a state passing through} a random-unitary channel of the form \changed{in Eq.~\eqref{eq:randomunitary}}, $\rho \rightarrow \tilde{\rho} = \mathcal{E}(\rho)$, the expectation value of an observable $\hat{O}$ is
\begin{subequations}
\begin{align}
    \expval{\hat{O}}_{\tilde{\rho}} &= \operatorname{Tr} \left[ \hat{O} \tilde{\rho} \right] \\ 
    &= \sum_i p_i \operatorname{Tr} \left[ \hat{O} U_i \rho U_i^{\dagger}\right]. \label{eq:wsum}
\end{align}
\end{subequations}
Typically, ancilla-based methods attempt to faithfully produce $\tilde{\rho}$, which can \changed{then} be accessed directly via measurement.


In our approach, we obtain the above expectation via an estimator of the random-unitary channel, following the formalism of Arrasmith \textit{et al}. \cite{arrasmith_operator_2020}. \changed{Given the probability distribution $\{p_i\}$ and $N$ total shots, we generate the following multinomial distribution:
\begin{equation}
S \sim \operatorname{Mult}(N,\{p_i\}) = (\hat{s}_1,\hat{s}_2, \ldots \hat{s}_k ).
\end{equation}
Here, the} $\hat{s}_i$ outcomes correspond to quantum circuits measuring the operator $U_i^{} \hat{O} U^\dagger_i $. Then, the estimator of the expectation values has the form
\begin{equation}\label{estimator}
    \hat{E} = \frac{1}{N} \sum_i \hat{s}_i \operatorname{Tr} \left[ \hat{O} U_i^{} \rho U^\dagger_i  \right],
\end{equation}
where $\mathbb{E}[\hat{s}_i] = p_i N$. This matches the form of Eq.~\eqref{eq:wsum} and is an unbiased estimator. Figure \ref{big_fig} represents this scheme pictorially. \changed{Because the probability distribution $\{p_i\}$ naturally sums to one, the variance of the estimator is simply the variance of the observable $\hat{O}$ divided by the number of shots $N$. Notably, it follows that the complexity and specific parameters $\{p_i\}$ of the random-unitary channel have no effect on the precision of the estimator.}

We note that a similar heuristic approach was tried for low-dimensional systems, but associated with exponential scaling problems \cite{rost_simulation_2020} and was limited to single-qubit applications \cite{tolunayHamiltonianSimulationQuantum2023}. Crucially, this work demonstrates the feasibility and scalability of this method even for exponentially \changed{complex} probability distributions $\{p_i\}$, as long as this distribution can be sampled efficiently. We demonstrate these ideas for simple open systems which, even in lieu of error mitigation, show constant \changed{precision over exponentially scaling problems.}

\section{Demonstrations of Scalable Random-Unitary Simulations} \label{depol_results}

To highlight our approach, we present two cases, both of which lead to exponentially scaling problems. We first look at examples of the depolarizing channel, a Pauli channel \changed{with exponentially many operators with respect to the number of qubits}. We then look at a discretized time-evolution process, \changed{whose operator complexity} grows exponentially in time. In the first instance, due to the simplicity of the method and the lack of ancilla costs required, we are able to easily demonstrate this approach on superconducting transmon qubit devices. The second example allows for a straightforward demonstration on quantum devices as well, although it can also be classically simulated via exact storage of the two-qubit density matrix. 
\subsection{Depolarizing channel}

Let $P_i^{(n)}$ denote an $n$-qubit Pauli string, i.e. $P_i^{(n)} \in\{I, X, Y, Z\}^{\otimes n}$. Then the $n$-qubit depolarizing channel \cite{nielsen_quantum_2010} can be written as
\begin{equation} \label{depolarizing_channel}
\mathcal{E}(\rho)=(1-p) \rho+\frac{p}{4^n-1} \sum_{i=1}^{4^n-1} P_i^{(n)} \rho P_i^{(n)}.
\end{equation}
For $n \in [1,27]$ (with ibmq\_montreal having $27$ qubits), we simulate the depolarizing channel on ibmq\_montreal with strength parameter $p = 0.5$ and an initial state of $|0^n \rangle$. For $n \in [1,3]$, we also demonstrate an ancilla-based approach, using a simple linear combination of unitaries. While many such methods exist, we compare simply against the most common example \cite{childs_hamiltonian_2012}, as generally all of these methods require ancilla circuits. \changed{The resource requirements} of the two methods are compared in Table \ref{table:resources}. We measure the set of single-Z observables $\langle Z^{(i)} \rangle$, e.g., $Z \otimes I$ for two qubits or $I \otimes Z \otimes I$ for three, and compare to the following analytic result:
\begin{equation} \label{z_analytic}
  \langle \hat{Z}^{(i)} \rangle = 1-\left(\frac{4^n}{4^n-1}\right) p.  
\end{equation}
We plot the mean squared error, ${\rm MSE} = \frac{1}{n} \sum_i^n (\langle Z^{(i)} \rangle - \langle \hat{Z}^{(i)} \rangle)^2$, against the number of qubits for both the ancilla and low-depth methods.

\begin{figure}[t!]
\includegraphics[width=\columnwidth]{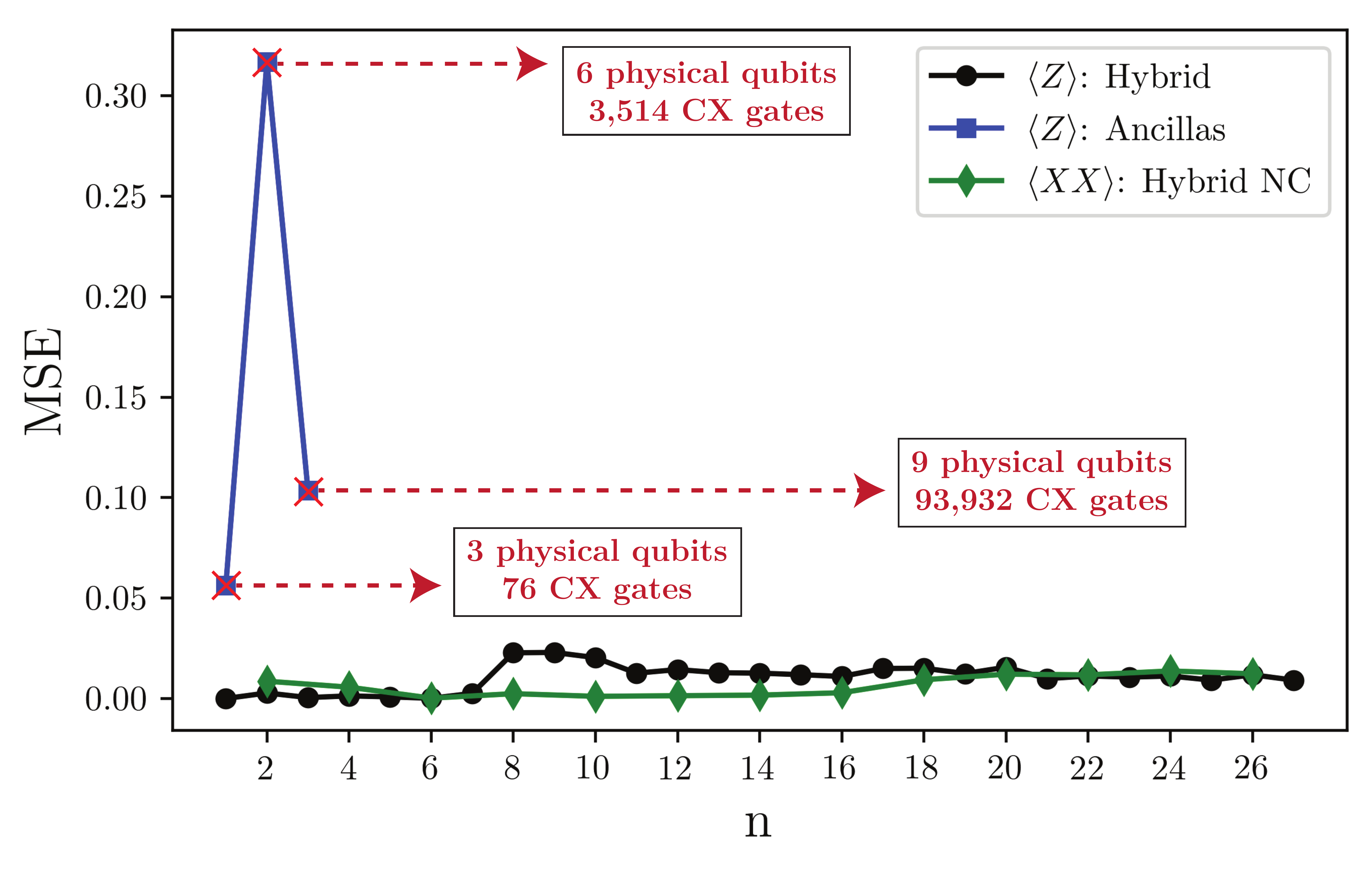} 
\caption{Simulation of the $n$-qubit depolarizing channel on the ibmq\_montreal device. The black and blue curves use the starting \changed{state $\left|0^n\right\rangle$ and} compare the ancilla-based approach to the proposed hybrid sampling method. As suggested by the red $\times$ marks, the ancilla data is shown purely to contrast with the hybrid method. Based on the average errors, the number of CNOTs, even for $n=1$, would result in a highly decohered state, so the specific MSE values are not particularly meaningful. The green curve performs the same hybrid simulation but with each pair of qubits prepared in the non-Clifford (NC) starting state $\frac{1}{\sqrt{2}} \ket{00} + \frac{e^{i \pi / 4}}{\sqrt{2}} \ket{11}$, and with measurements in the $X$ basis. Each point corresponds to $10^3$ shots.}
\label{depol_figure}
\end{figure} 

\changed{Despite the depolarizing channel's exponentially large sample space of $4^n$ operators, the probability distribution $\{p_i\}$ can be sampled efficiently. Because these operations belong to the Clifford group \cite{gottesman_theory_1998}, this channel can be classically simulated in polynomial time according to the Gottesman-Knill theorem \cite{gottesman_heisenberg_1998}. Thus, we} also include an instance of a simple non-Clifford state, where we prepare the qubits pairwise in the entangled state $\frac{1}{\sqrt{2}} \ket{00} + \frac{e^{i \pi / 4}}{\sqrt{2}} \ket{11}$ and simulate the depolarizing channel for even values of $n$. We measure the set of pairwise observables $\left\langle\hat{X}^{(2i)}\hat{X}^{(2i+1)}\right\rangle$ and compare to the following analytic result:
\begin{equation}\label{xx_analytic}
  \left\langle\hat{X}^{(2i)}\hat{X}^{(2i+1)}\right\rangle=\frac{1}{\sqrt{2}}\left(1-\left(\frac{4^n}{4^n-1}\right) p\right) .
\end{equation}
We plot the mean squared error against the number of simulated qubits using our low-depth hybrid method. All results are combined in Fig. \ref{depol_figure}. As shown, our hybrid sampling method clearly outperforms the traditional ancilla-based approach. It also demonstrates consistently reliable results up to the maximum number of qubits. In contrast, circuits to run the multiplexing operations for the ancilla-based method become infeasible after 3 qubits.

\begin{figure}[t!]
\includegraphics[width=\columnwidth]{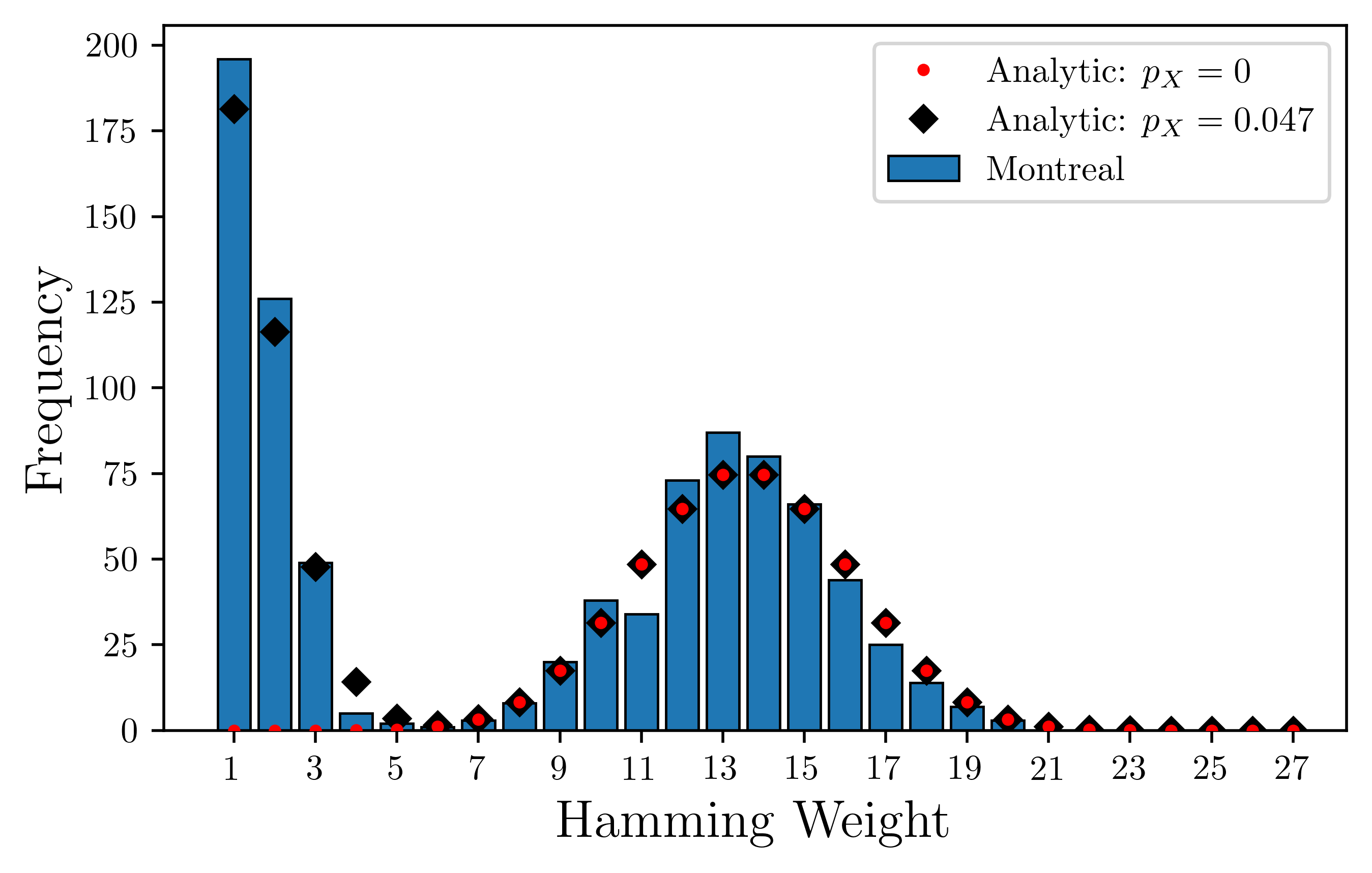} 
\caption{Hamming weight distribution for the 27-qubit depolarizing channel, with starting state $\left|0^n\right\rangle$. From the 1000 total measurements, we plot the frequency of each Hamming weight. The blue bars are data from the ibmq\_montreal device, the red dots are analytically calculated ideal results, and the black diamonds are analytically calculated results that account for bit-flip errors. The bit-flip effects are most visible at the start of the distribution, as many bit-flip errors symmetrically cancel around the center.}
\label{hamming_figure}
\end{figure}

Despite the high numbers of qubits and the prevalence of errors, the simulations here still manage to faithfully capture the simulated state dynamics. In Fig. \ref{hamming_figure}, we plot the Hamming weight distribution of the 27-qubit measurements alongside analytic expectation values. Expected bit-flip errors across each qubit skew the results away from the \changed{all-zero} state, yet not in a way that corrupts the overall distribution. We model the overall distribution as the depolarizing channel coupled with bit-flip errors of $p_X=4.7\%$ per qubit.

\renewcommand{\arraystretch}{1.5}
\begin{table}[]
    \centering
    \begin{tabular}{c|c|c}
        \multirow{2}{*}{{\bf Method} }  &   \textbf{Stinespring} &  \textbf{Hybrid}   \\[-5pt]
        & {\bf dilation} & {\bf sampling} \\ \hline 
      {\bf Qubits}  & $n+\lceil\log (m)\rceil $ & $n$ \\
      {\bf Depth} & $\Omega(m)$ & $1$ \\
      {\bf Circuits} & $1$ & $\min (m, N)$ \\
      {\bf Variance} & $\frac{1}{N}\left(\left\langle O^2\right\rangle_{\tilde{\rho}} - \left\langle O\right\rangle^2_{\tilde{\rho}}  \right)$  & $\frac{1}{N}\left(\sum_i p_i\left\langle O^2\right\rangle_{\tilde{\rho_i}} - \langle O \rangle_{\tilde{\rho}}^2\right)$ 
    \end{tabular}
    \caption{Comparison of resources used in simulating $n$-qubit quantum channels, comprised of a convex sum of $m$ unitary evolutions.}
    \label{table:resources}
\end{table}

\subsection{Noisy Hamiltonian evolution} \label{section:tfim}
To look at the performance under a different exponentially \changed{growing} scheme, we simulate the two-qubit transverse-field Ising model (TFIM) under a noise channel which occurs at each discretized step. 

\begin{figure}[t!]
\begin{center}
\includegraphics[width=\columnwidth]{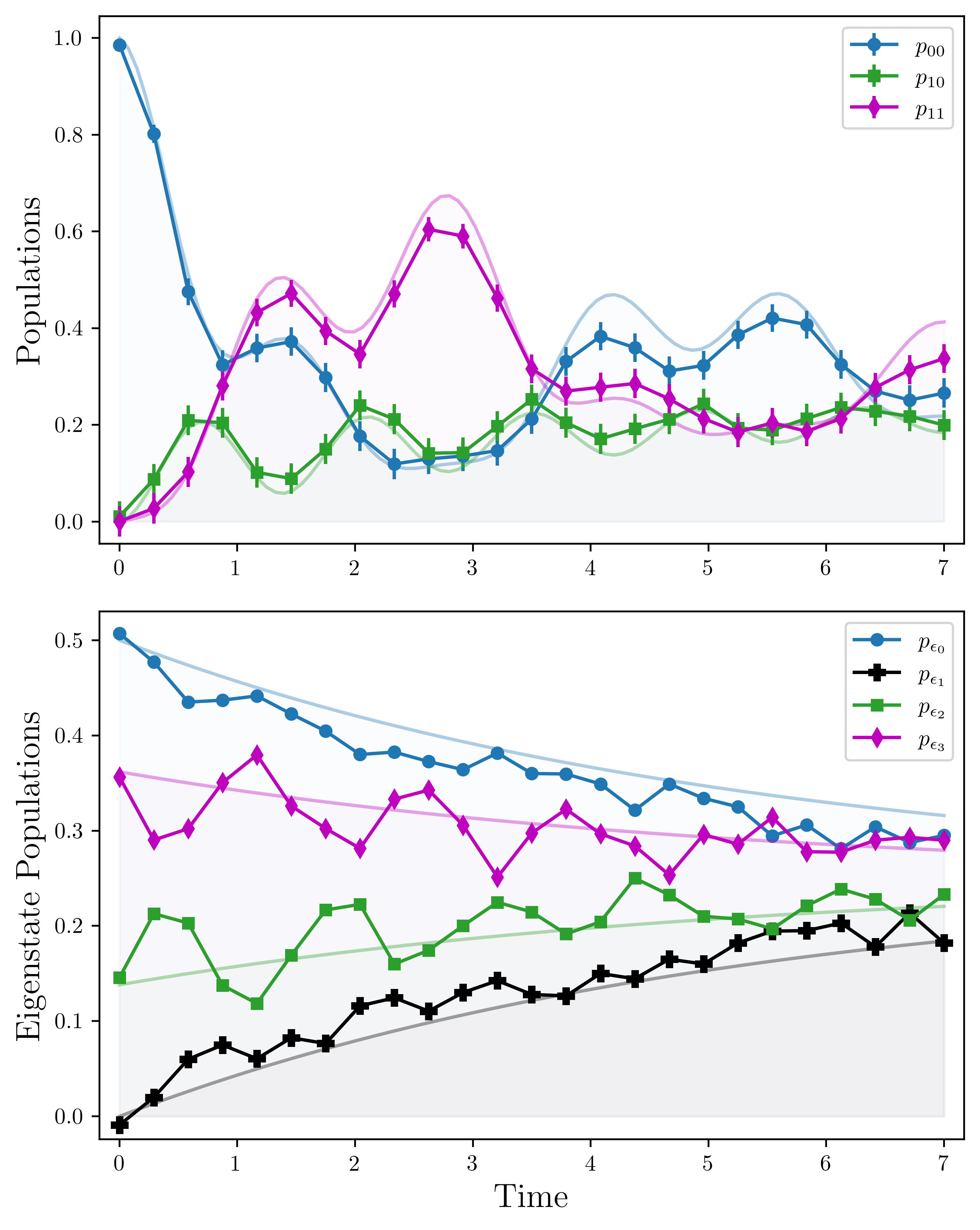} 
\caption{Simulation of the two-qubit TFIM subject to depolarizing noise. The points represent data from ibmq\_kolkata, and the smooth curves represent the analytic results. The top figure displays populations over time in the standard basis, while the bottom figure displays populations in the eigenbasis of the TFIM Hamiltonian. Note that analytically $p_{01}(t) = p_{10}(t)$, so we omit $p_{01}(t)$ for visual clarity.}
\label{tfim_figure}
\end{center}
\end{figure}

The TFIM Hamiltonian is 
\begin{equation}
H=-J \sigma_z^{(1)} \sigma_z^{(2)}-h\left(\sigma_x^{(1)}+\sigma_x^{(2)}\right),
\end{equation}
where $J$ is the exchange interaction parameter and $h$ quantifies the strength of the transverse magnetic field. For small time steps $\Delta t$, we evolve this system as the Hamiltonian \changed{evolution operator composed with the random-unitary channel $\mathcal{S}$}. The resulting composed operator is itself a random-unitary channel, denoted as $\tilde{\mathcal{S}}$. We generate each time step recursively, by sampling a distribution as in Eq.~\eqref{estimator}:
\begin{equation}
\tilde{\mathcal{S}}_{t+1} = \tilde{\mathcal{S} }\circ \tilde{\mathcal{S}}_{t}.
\end{equation}
We simulate these dynamics on the ibmq\_kolkata device, with a depolarizing strength of $p=0.05$ per time step, \changed{TFIM parameters $J=1$ and $h=1$, and the starting state $\left|00\right\rangle$}. Figure \ref{tfim_figure} shows the populations over time, both in the standard computational basis and in the eigenbasis of the Hamiltonian. \changed{Absent any noise, the eigenstates would be stabilized at constant populations}. Fitting exponential decay functions to the eigenstate population curves yields an estimated decoherence time of $T_1 = 4.84 \pm 0.46$ (a.u.), encompassing the predicted time of $T_1 = 5.25$ (a.u.).

In this example, the possible number of circuits to sample from grows exponentially with the number of time steps. For our simulations, after 25 time steps there are $16^{25}$ circuit permutations to consider, a seemingly intractable sample space. However, due to the constant norm of the probability distribution vector, the measurement statistics are agnostic to the complexity of the sample space. This enables our results to accurately capture the system-environment dynamics while using only $10^3$ shots at each \changed{point in time.}

\section{Discussion and Conclusion}

These results demonstrate that for a certain class of quantum channels, \changed{the proposed hybrid quantum-classical approach offers significant efficiency advantages. By stochastically generating low-depth quantum circuits and then aggregating the measurement results, we enable the simulation of exponentially complex processes.}

\changed{A key result in our work is that the variance of our estimator is unchanged by the specifics of the probability distribution. Because random-unitary channels consist of a convex combination of unitary operations, the set of sampling probabilities $\{p_i\}$ always sums to unity. While we chose the depolarizing channel to demonstrate scalability, any random-unitary channel can be simulated via this method, assuming only that its probability distribution can be sampled efficiently. This includes non uniform noise channels, in which each unitary operator $U_i$ has a distinct probability $p_i$.}

\changed{While our method leads to efficiency gains important for noisy quantum device applications, one limitation is that it does not create} a coherent quantum state as an output. This means that any observables obtained via further evolutions of the state must also be processed through an estimator. For instance, if one wishes to use a mixed state as input to some broader algorithm, then any desired observables at the end of this algorithm must themselves be sampled using an estimator of the form in \changed{Eq.~\eqref{estimator}. However, if the channel's operator representation has exponentially many terms, implementation via ancilla-based methods will be significantly challenging as well}. This problem is demonstrated in Fig. \ref{depol_figure}, where simulating \changed{a 3-qubit depolarizing channel with a linear dilation technique naively} requires nearly $10^5$ CX gates.

Other difficulties exist in practically implementing these simulations on quantum computers. For instance, random distributions that cannot be sampled in constant or polynomial time could themselves result in memory issues. Another problem exists for $U_i$ that cannot be efficiently implemented as products of two-qubit unitaries, i.e. unitaries with exponential complexity in their generators, although this does not exclude their generation via more sophisticated quantum simulators. These problems would also plague ancilla-based approaches.

Our method \changed{is related to other} stochastic techniques within quantum simulation \changed{and open quantum systems simulation, where random effects can help reduce circuit requirements or classical run times} \cite{wallman_noise_2016, moueddene_realistic_2020, suzuki_qulacs_2021, terhal_quantum_2015, temme_error_2017, rost_simulation_2020, tolunayHamiltonianSimulationQuantum2023}. Stochastic compilation methods have also notably been used to construct first-order dynamics of Hamiltonian evolution \cite{campbell_random_2019} and twirled channels \cite{kimScalableErrorMitigation2023}, and as a tool in implementing certain shallow channels \cite{rost_simulation_2020}. 

\changed{While random-unitary channels do not encompass all open system dynamics, they are an important subset of quantum channels, and we envision applications of this protocol beyond the examples shown in this work. For example, noise on quantum computers can be modeled as Pauli channels, with empirically determined strength parameters $p_i$ \mbox{\cite{moueddene_realistic_2020}}. Further, more general quantum channels can be simulated using approximate random unitary forms \cite{rosgenAdditivityDistinguishabilityRandom2008}.} 

Modeling environmental effects is essential in understanding the dynamics of many physical systems. The current work proposes a straightforward way to simulate random-unitary noise channels via classical probabilistic sampling of low-depth unitaries. For quantum simulations of random-unitary processes, the proposed hybrid approach has clear advantages over ancilla-based techniques. While the scope is limited to random-unitary channels, these represent an important class of quantum channels \changed{and open} the door for a variety of other dynamics to be simulated on near-term quantum devices. 

\section{Acknowledgments}
This work is supported by the NSF RAISE-QAC-QSA, Grant No. DMR-2037783; the Department of Energy, Office of Basic Energy Sciences, Grant No. DE-SC0019215; the NSF DGE NRT-QISE, Grant No. 2125924; the NSF ECCS CAREER, Grant No. 1944085; and the NSF CNS, Grant No. 2247007.

The authors acknowledge the use of IBM Quantum services for this work. The views expressed are those of the authors and do not reflect the official policy or position of IBM or the IBM Quantum team.

\section{Appendix} \label{section: appendix}
\subsection{Shot allocation}
While the hybrid simulation method laid out in Sec. \ref{section: intro} is robust to different sampling schemes, we optimize the use of valuable computing resources with shot-frugal allocation methods. Practically, some of the operators within a random-unitary channel demand higher priority due to their different weightings $p_i$ in \changed{Eq.~\eqref{eq:wsum}}. Throughout these results, we use weighted random sampling of the probability distributions $\{p_i\}$ to allocate shots, thus optimizing the use of computing time \cite{arrasmith_operator_2020}. More precisely, we sample the operator space $\{U_i\}$ with corresponding probabilities $\{p_i\}$ to determine the circuit run for each shot. This means that for a simulation using a total of $N$ shots, on average $p_i N$ shots will be allocated to each operator $U_i$ within the sampled space.

To compute the variance of our estimator in \changed{Eq.~\eqref{estimator}}, we follow the weighted random sampling calculations of Arrasmith \textit{et al}. \cite{arrasmith_operator_2020}. This gives the following variance:
\begin{equation} \label{hybrid_variance}
\operatorname{Var}(\hat{E})=\frac{1}{N}\left(\sum_i p_i\left\langle O^2\right\rangle_{\tilde{\rho_i}} - \langle O \rangle_{\tilde{\rho}}^2\right),
\end{equation}
where $\tilde{\rho_i} := U_i \rho U_i^{\dagger}$. In comparison, when using Stinespring dilation, the variance of the estimator $\hat{E}_S$ is simply $\frac{1}{N}$ times the variance of the expectation value $\expval{O}_{\tilde{\rho}}$:
\begin{equation}
\operatorname{Var}(\hat{E}_S) = \frac{1}{N} \left(\left\langle O^2\right\rangle_{\tilde{\rho}} - \left\langle O\right\rangle^2_{\tilde{\rho}}  \right).
\end{equation}
Analytically, these expressions are equivalent: Applying \changed{Eq.~\eqref{eq:wsum}} to $O^2$ yields $\expval{O^2}_{\tilde{\rho}} = \sum_i p_i\left\langle O^2\right\rangle_{\tilde{\rho_i}}$. However, our hybrid sampling can only provide access to the individual expectation values $\left\langle O^2\right\rangle_{\tilde{\rho_i}}$, making \changed{Eq.~\eqref{hybrid_variance}} the practical means of calculating the variance.

\begin{figure}[t!]
\centering
\includegraphics[scale=0.185]{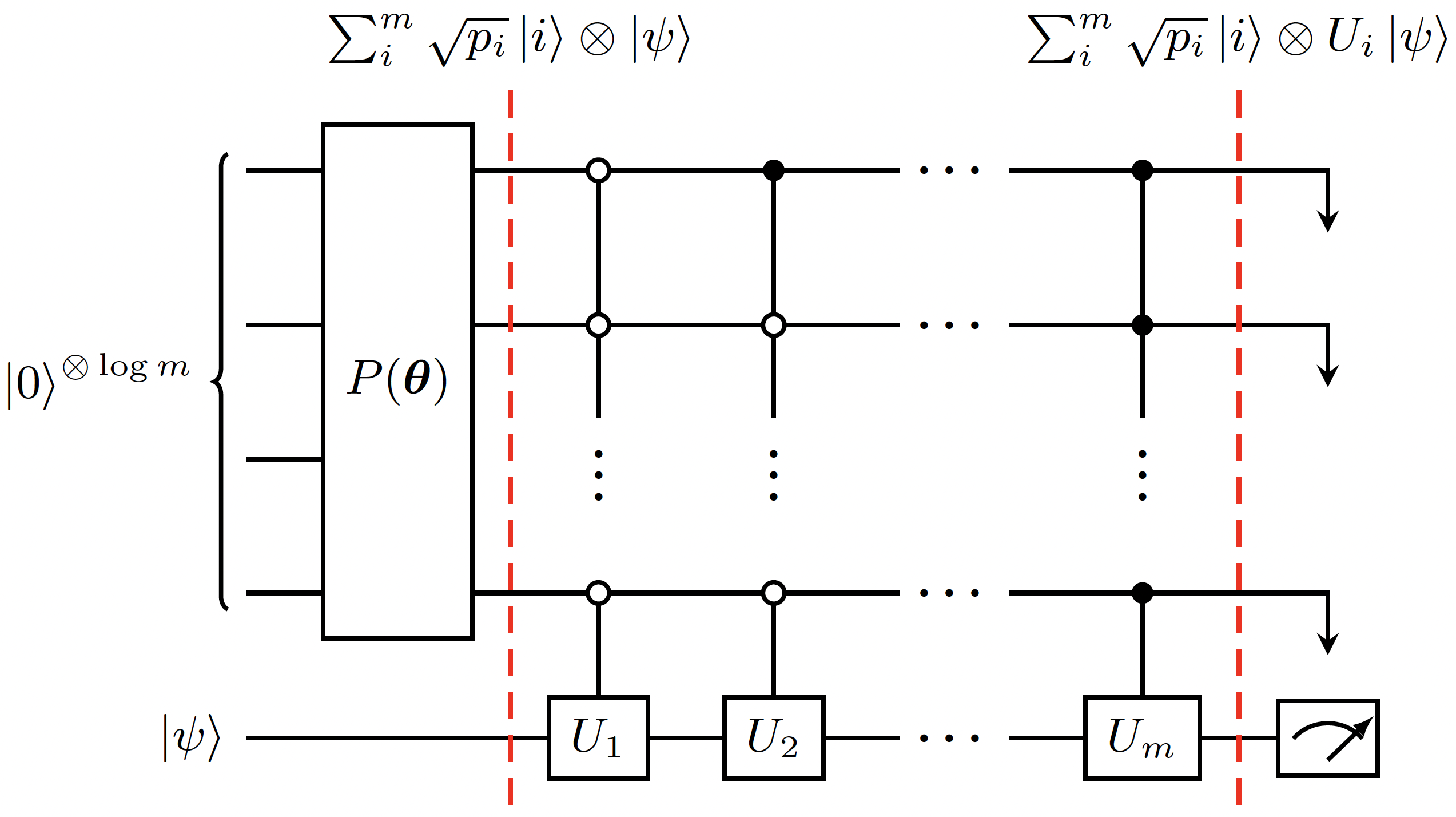}

\caption{Circuit schematic for ancilla-based simulation method \cite{childs_hamiltonian_2012}. The state of the combined ancilla-target system is displayed above each red, dotted line.}
\label{fig:ancillas}
\end{figure}

\subsection{Analytic expectation values}
The $n$-qubit depolarizing channel in \changed{Eq.~\eqref{depolarizing_channel}} can equivalently be written as follows \cite{nielsen_quantum_2010}:
$$\mathcal{E}(\rho)=(1-\lambda) \rho + \lambda\frac{I}{2^n},$$
where $\lambda = (\frac{4^n}{4^n-1}) p$. This form greatly simplifies analytic calculations, and, in fact, the expectation value of a general observable $\hat{O}$ evolved under $\mathcal{E}$ can be calculated as follows:
\begin{equation}
    \expval{\hat{O}}_{\tilde{\rho}} = (1-\lambda) \operatorname{Tr}\left[O  \rho \right] + \frac{\lambda}{2^n} \operatorname{Tr}\left[O \right].
\end{equation}
From this form, we computed the analytic expectation values $\langle \hat{Z}^{(i)} \rangle$ and $\left\langle\hat{X}^{(2i)}\hat{X}^{(2i+1)}\right\rangle$ in \changed{Eqs. \eqref{z_analytic} and \eqref{xx_analytic}}, respectively.

\subsection{Ancilla-based algorithm}
For the depolarizing channel simulations, we compare with results from an ancilla-based simulation algorithm \cite{childs_hamiltonian_2012}, described as follows. Consider the random-unitary channel $\mathcal{E}(\rho)=\sum_i^m p_i\left(U_i \rho U_i^{\dagger}\right)$. First, encode the probabilities $p_i$ as amplitudes of $\lceil\log (\mathrm{m})\rceil$-many ancilla qubits. Suppose a gate $P(\bm{\theta})$ exists such that:

\begin{equation}
    P(\bm{\theta}) \ket{0^{\lceil\log (\mathrm{m})\rceil}} = \sum_i^m \sqrt{p_i} \ket{i}.
\end{equation}

Then, apply each unitary evolution $U_i$ to the target state $\ket{\psi}$ conditioned on the ancilla state $\ket{i}$. This accomplishes the desired effect of applying each unitary $U_i$ with probability $p_i$. The overall algorithm is displayed in circuit form in Fig. \ref{fig:ancillas}.

%


\end{document}